 \documentclass[preprint,review,12pt]{elsarticle}

 \usepackage{graphicx}

\usepackage{amssymb}
\usepackage{epstopdf}
\usepackage{latexsym}

\journal{Physics Letters B}

\begin{document}

\begin{frontmatter}



\title{Produced charged hadrons in central Pb + Pb collisions at LHC energies in the RDM}


\author{Georg Wolschin}

\address{ Institut f{\"ur} Theoretische 
Physik
der Universit{\"a}t Heidelberg, 
        Philosophenweg 16,  
        D-69120 Heidelberg, Germany}

\begin{abstract}
The energy dependence of charged-hadron production in relativistic heavy-ion 
collisions is investigated in a nonequilibrium-statistical relativistic diffusion model (RDM) 
with three sources. Theoretical pseudorapidity distributions are compared with 
Au + Au data at RHIC energies of $\sqrt {s_{NN}}$ = 0.13 and 0.2 TeV, and computed for Pb + Pb central collisions at LHC energies of 2.76 and 5.52 TeV. The nearly equilibrated source at midrapidity arising from gluon-gluon collisions becomes the major origin of particle production at LHC energies. The midrapidity dip is determined by the interplay of the three sources.\\
\end{abstract}

\begin{keyword}
Relativistic heavy-ion collisions \sep Charged-hadron pseudorapidity distributions \sep Relativistic Diffusion Model \sep Predictions at LHC energies 
\PACS 25.75.-q \sep 25.75.Dw \sep 12.38.Mh

\end{keyword}

\end{frontmatter}



\newpage
\section{\label{sec:intro}Introduction\protect} 

With the advent of first results from heavy-ion collisions at LHC energies of $\sqrt {s_{NN}}$ = 2.76 TeV in central Pb + Pb collisions \cite{aa10,aamo10}, a new perspective on this area of research opens up. The strong gluon field that is present at these high energies determines the dynamics of the collision and the details of particle production even more decisively than in Au + Au collisions at RHIC energies of 0.13 and 0.2 TeV, where quark-gluon interactions are still more important in the particle production process than gluon-gluon collisions.
The first and simplest observable to be determined experimentally is the charged-particle multiplicity density at mid-rapidity in central Pb + Pb. There are many theoretical models predicting this value with varying accuracy (see \cite{arm08,arme09} at the maximum LHC energy of 5.52 TeV, and \cite{aa10} at 2.76 TeV). However, the experimental ALICE result 
of 1601 $\pm 60$ \cite{aamo10} at 2.76 TeV is obtained from a straightforward extrapolation of the midrapidity values at RHIC energies with $\log(\sqrt {s_{NN}}$). 

More specific information can be expected from the detailed shape of the pseudorapidity distribution of produced charged hadrons at $\eta-$values further away from midrapidity, which will be available experimentally in the near future. The decomposition of the distribution function $(dN/d\eta)(\eta)$ from the underlying physical ingredients such as quark-gluon vs. gluon-gluon interactions will be of particular interest. 

In this Letter an analytically soluble nonequilibrium-statistical RDM-model \cite{wol99,wol07} that successfully describes pseudorapidity distributions for produced hadrons at RHIC energies is used to predict these distribution functions at LHC energies. The model relies on three sources for charged-hadron production, with the midrapidity source associated with gluon-gluon collisions, and two forward-centered fragmentation sources arising essentially from valence quark -- gluon interactions. 


It has been shown in \cite{biy04,wob06,kw07} within the relativistic diffusion model (RDM) that at RHIC energies of 0.13 TeV (0.2 TeV) the midrapidity source generates about  13 \% (26 \%) of the produced particles in a 0--6\% central Au + Au collision, whereas the bulk of the particles is still produced in the two fragmentation sources. At SPS, and low RHIC energies of 19.6 GeV the effect of the midrapidity source is  negligible \cite{kw07}.

In the asymmetric $d$ + Au system at 0.2 TeV there is also a sizeable midrapidity source containing 19 \% of the produced particles for 0--20\% central collisions \cite{wobi06}. Particle creation from a gluon-dominated midrapidity source, incoherently added to the sources related to the valence part of the nucleons, had also been proposed by Bialas and Czyz \cite{bia05}. 
There exist also many other models which assume a central source such as the dual parton model \cite{cap82,cap94}, or the quark-gluon string model \cite{kai03}. The RDM provides an analytical framework to investigate the interplay of central and fragmentation sources.

For asymmetric systems, the central source is shifting in rapidity space with increasing centrality, whereas for symmetric systems it remains at midrapidity $<\eta>=0$. The shape of the $dN/d\eta$-distributions at different centralities is very sensitive to the detailed balance of the underlying distribution functions, and the excellent agreement with the $d$ + Au PHOBOS-data \cite{bb04,bb05,alv11} at 0.2 TeV lends credibility to the three-sources model also for symmetric systems where the details of the distribution functions are less specific.

Within the RDM, I investigate in this Letter the energy dependence of the three sources for particle production in central collisions of symmetric systems, and provide predictions at LHC energies. The energy range considered here for the three-sources model covers RHIC energies of $\sqrt {s_{NN}}$ = 0.13 and 0.2 TeV in Au + Au collisions, the presently accessible LHC energy of 2.76 TeV in Pb + Pb collisions, and the maximum LHC energy of 5.52 TeV.

The model is considered in Sec. 2, the calculation of pseudorapidity distributions of charged hadrons at RHIC and LHC energies in Sec. 3, and conclusions are drawn in Sec. 4.

\section{Relativistic Diffusion Model}

\begin{figure}
\begin{center}%
\includegraphics[width=12cm]{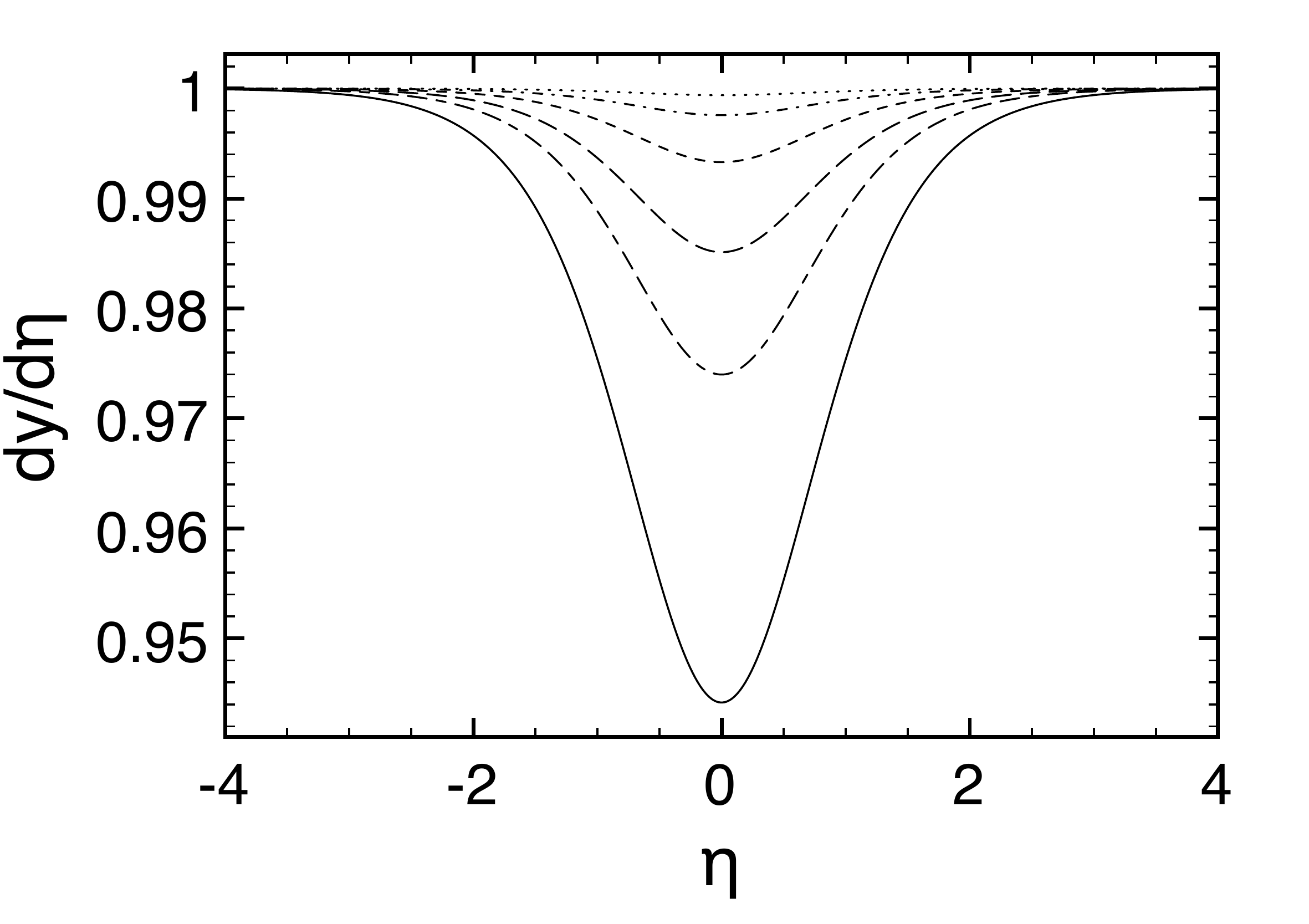}
\caption{\label{fig1} The Jacobian $dy/d\eta$ for $<m> = m_{\pi}$ and average transverse momenta (bottom to top) $<p_T> = 0.4, 0.6, 0.8, 1.2, 2$ and 4 GeV/c.}
\end{center}
\end{figure}

In the Relativistic Diffusion Model, the
rapidity distribution of produced particles
emerges from an incoherent superposition of the beam-like fragmentation 
components at larger rapidities arising mostly from valence quark-gluon interactions, and a
component centered at midrapidity that is essentially
due to gluon-gluon collisions. All three distributions are broadened in rapidity space
as a consequence of diffusion-like processes.


The time evolution of the distribution
functions is governed by a Fokker-Planck
equation (FPE) in rapidity space
\cite{wol07} (and references therein)
\begin{equation}
\frac{\partial}{\partial t}[ R(y,t)]^{\mu}=-\frac{\partial}
{\partial y}\Bigl[J(y)[R(y,t)]^{\mu}\Bigr]+
\frac{\partial^2}{\partial y^2}[D_{y}\cdot R(y,t)]^{\nu} 
\label{fpenl}
\end{equation}
with the rapidity $y=0.5\cdot \ln((E+p)/(E-p))$. The beam rapidity can also be written as 
$y_{beam}=\mp y_{max}=\mp \ln(\sqrt{s_{NN}}/m_{p})$.
The rapidity diffusion coefficient $D_{y}$ that contains the
microscopic physics accounts for the broadening of the
rapidity distributions.
The drift $J(y)$ determines the shift of the mean rapidities
towards the central value, and linear and nonlinear 
forms have been discussed \cite{alb00,ryb03,wol07}.

The standard linear FPE corresponds to $\mu=\nu=1$ and a
 linear drift function
 \begin{equation}
J(y)=(y_{eq}- y)/\tau_{y}
\label{dri}
\end{equation}
with the rapidity relaxation time $\tau_{y}$, and the equilibrium 
value $y_{eq}$ of the rapidity. This is
the so-called Uhlenbeck-Ornstein \cite{uhl30} process, applied to the
relativistic invariant rapidity for the three components  
$R_{k}(y,t)$ ($k$=1,2,3) of the distribution function
in rapidity space
\begin{equation}
\frac{\partial}{\partial t}R_{k}(y,t)=
-\frac{1}{\tau_{y}}\frac{\partial}
{\partial y}\Bigl[(y_{eq}-y)\cdot R_{k}(y,t)\Bigr]
+\frac{\partial^2}{\partial y^2}
\Bigl[ D_{y}^{k}\cdot R_{k}(y,t)\Bigr].
\label{fpe}
\end{equation}

Since the equation is linear, a superposition of the distribution
functions \cite{wol99,wol03} using the initial conditions
$R_{1,2}(y,t=0)=\delta(y\pm y_{max})$
with the absolute value of the beam rapidities 
$y_{max}$, and $R_{3}(y,t=0)=\delta(y-y_{eq})$  
yields the exact solution. In the solution, the mean values 
are obtained analytically from the moments 
equations as
\begin{equation}
<y_{1,2}(t)>=y_{eq}[1-exp(-t/\tau_{y})] \mp y_{max}\exp{(-t/\tau_{y})}
\label{mean}
\end{equation}
for the sources (1) and (2) with the absolute value of the beam rapidity $y_{max}$,
and $y_{eq}$ for the local equilibrium source which is equal to zero only for
symmetric systems. Hence, both mean values $<y_{1,2}>$ would attain y$_{eq}$ 
for t$\rightarrow \infty$, whereas for short times they remain between 
beam and equilibrium values. The variances are
\begin{equation}
\sigma_{1,2,eq}^{2}(t)=D_{y}^{1,2,eq}\tau_{y}[1-\exp(-2t/\tau_{y})],
\label{var}
\end{equation}
and the corresponding FWHM-values
are obtained from 
$\Gamma=\sqrt{8\ln2}\cdot \sigma$ since the partial distribution functions are Gaussians in rapidity space
(but not in pseudorapidity space).

The midrapidity source has mean value zero and hence, comes close to thermal equilibrium with respect to the variable rapidity during the interaction time
$\tau_{int}$. Note that the width approaches equilibrium twice as fast as the mean value. I use the notion $R_{eq}(y,t)$ for the
associated partial distribution function in y-space, with 
$N_{ch}^{eq}$ charged particles, cf. Table~\ref{tab1}. Full equilibrium as determined by the temperature would be reached for $\tau_{int}/\tau_y \gg 1$. The fragmentation sources do not reach $<y_{1,2}>=0$ during the interaction time and hence, remain far from thermal distributions in rapidity space, and do not fully equilibrate with the central source.

\section{Pseudorapidity distributions}
If particle identification is not available, one has to
convert the results to pseudorapidity, 
$\eta=-$ln[tan($\theta / 2)]$ with the scattering angle $\theta$.
The conversion from $y-$ to $\eta-$
space of the rapidity density
\begin{equation}
\frac{dN}{d\eta}=\frac{dN}{dy}\frac{dy}{d\eta}=\frac{p}{E}\frac{dN}{dy}\simeq
J(\eta,\langle m\rangle/\langle p_{T}\rangle)\frac{dN}{dy} 
\label{deta}
\end{equation}
is performed through the Jacobian
\begin{eqnarray}
\lefteqn{J(\eta,\langle m\rangle/\langle p_{T}\rangle)=\cosh({\eta})\cdot }
\nonumber\\&&
\qquad\qquad[1+(\langle m\rangle/\langle p_{T}\rangle)^{2}
+\sinh^{2}(\eta)]^{-1/2}.
\label{jac}
\end{eqnarray}

The average mass $<m>$ of produced charged hadrons in the
central region is approximated by the pion mass $m_{\pi}$ since pions represent by far the largest fraction of produced charged hadrons, in particular in the midrapidity source where the transformation has the
biggest effect. 

The dependence on the mean transverse momentum $<p_{T}>$ is illustrated in Fig.~\ref{fig1}.
Due to the Jacobian, the partial distribution functions differ from Gaussians. In the actual calculations, I use $<p_{T}>=0.3$ and 0.4 GeV at the respective RHIC energies of 0.13 and 0.2 TeV, and $<p_{T}>=0.6$ and 0.7 GeV at LHC energies of 2.76 and 5.52 TeV. The values at LHC energies should be updated once measured $p_T$-distributions become available.

\begin{figure}
\begin{center}
\includegraphics[width=8cm]{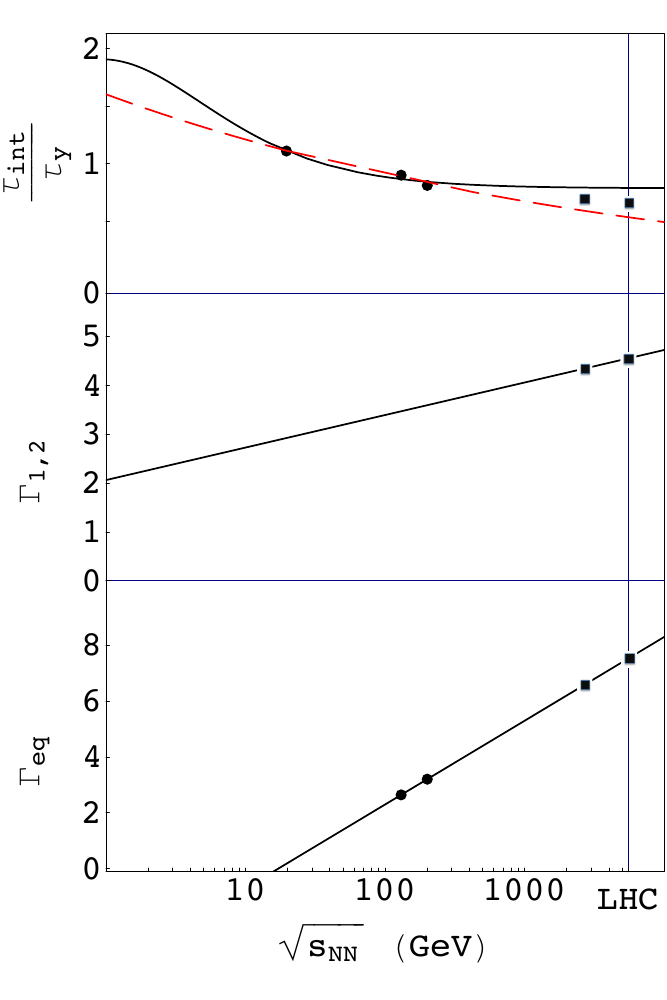}
\caption{\label{fig2}Dependence of the diffusion-model parameters 
for heavy systems
(central Au + Au at RHIC energies, central Pb + Pb at LHC energies) on the center-of-mass 
energy $\sqrt{s_{NN}}$ according to \cite{kw07}:  Quotient of interaction time and relaxation 
time for sinh- and exponential (dashed) extrapolation (upper frame, with rescaled absolute values);  
width of the peripheral sources including collective expansion (middle
frame); effective width of the midrapidity source (lower frame).
The results are for charged-hadron pseudorapidity distributions,
with extrapolations to LHC energies. The dots refer to the fit values at RHIC energies of 19.6, 130 and 200 GeV. The time parameters used in the present work at LHC energies of 2.76 and 5.52 TeV have been averaged between the two analytical extrapolations.}
\end{center}
\end{figure}

\begin{figure*}
\begin{center}
\includegraphics[width=12cm]{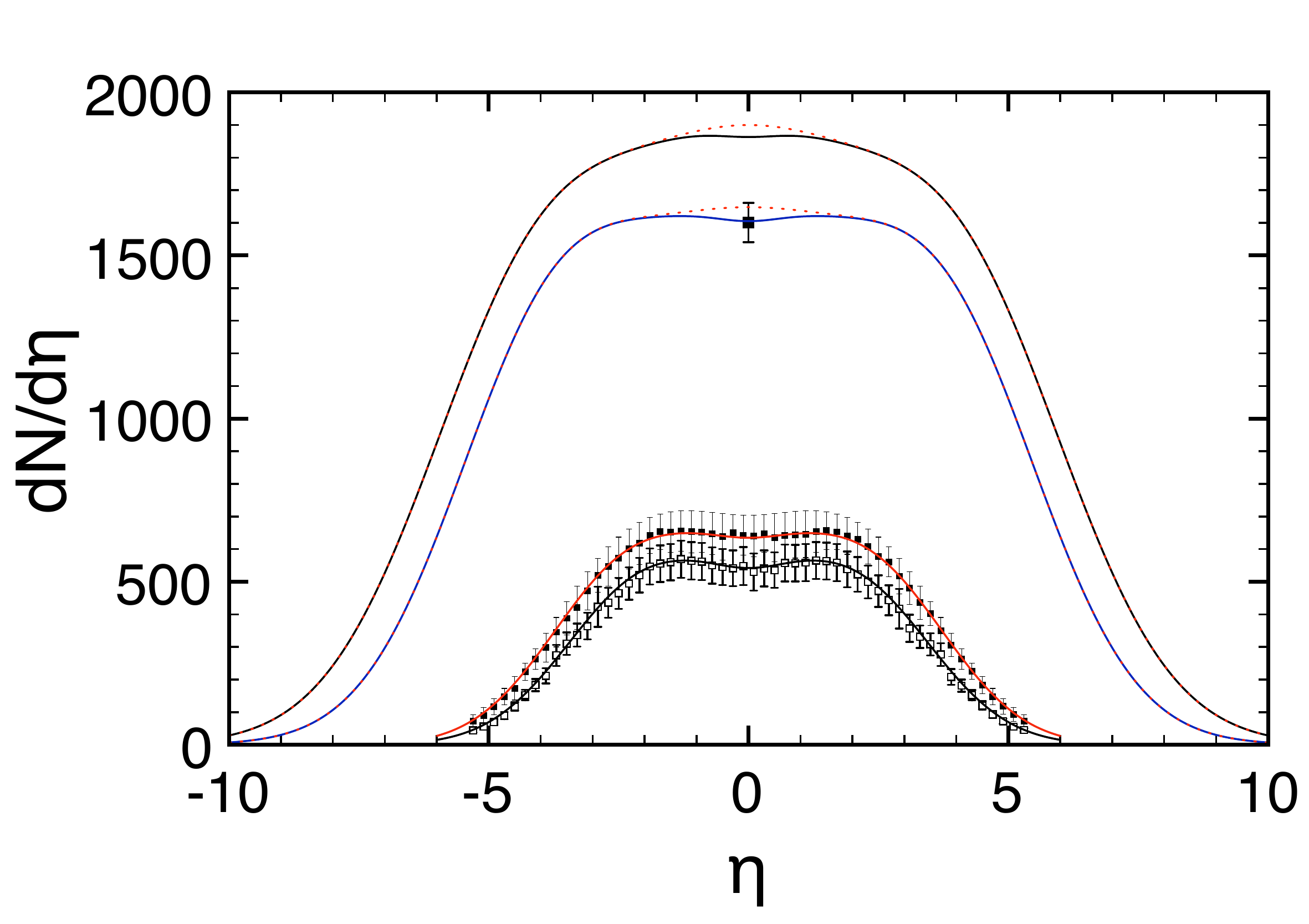}
\caption{\label{fig3}Calculated pseudorapidity distributions of 
produced charged particles from
Au + Au collisions (bottom) at $\sqrt{s_{NN}}$ =  0.13 and 0.2 TeV for
0--6\% central collisions in comparison with
PHOBOS data \cite{bb01,bb03}. The analytical three-sources RDM-solutions are
optimized in a fit to the data. Distribution functions for 0--5\% central Pb + Pb collisions
at LHC energies of 2.76 and 5.52 TeV are shown in the upper part of the figure, with the lower-energy result adjusted to the recent midrapidity ALICE data point \cite{aamo10}. Dotted curves are without the Jacobian transformation. The corresponding 
parameter values are given in Table~\ref{tab1}.}
\end{center}
\end{figure*}

\begin{figure*}
\begin{center}
\includegraphics[width=12cm]{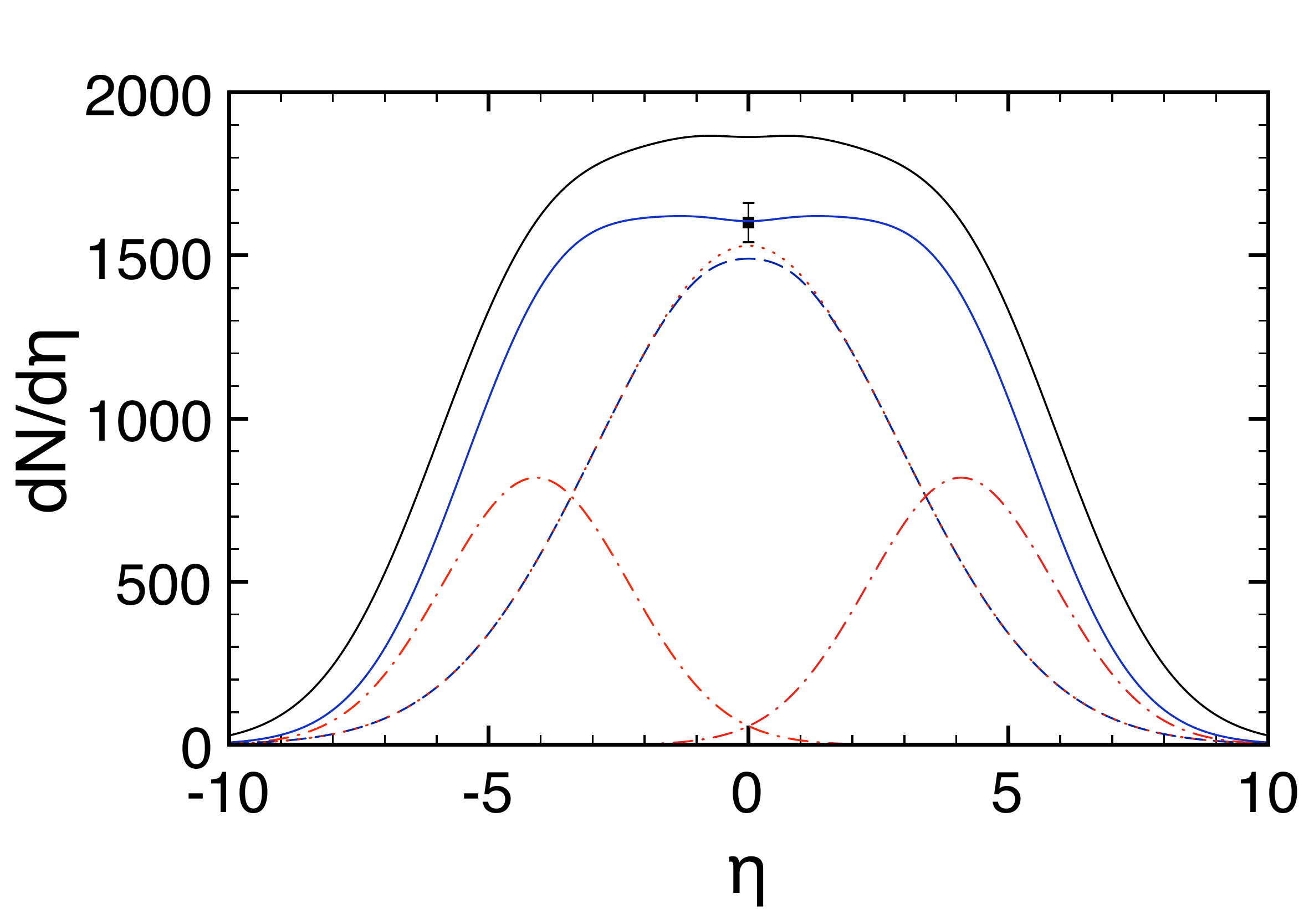}
\caption{\label{fig4}Pseudorapidity distributions of charged hadrons in 0--5\% central Pb + Pb collisions at LHC energies of \(\sqrt {s_{NN}}\) =  2.76 and 5.52 TeV. The underlying theoretical distributions are shown for 2.76 TeV. Their shapes are not significantly modified by the Jacobian.
The size of the midrapidity dip is determined by the interplay of central (gluon-gluon, dashed; without Jacobian, dotted) and peripheral (valence quarks -- gluon, dash-dotted) distribution functions. The midrapidity value is almost completely determined by particle production from gluon-gluon collisions at LHC energies.}
\end{center}
\end{figure*}

\begin{table}
\begin{center}
\caption{\label{tab1}Three-sources RDM-parameters for 0--6\% Au + Au at RHIC energies (upper two lines) and for 0--5\% Pb + Pb at LHC energies (lower two lines). See Fig.~\ref{fig2} and text for the extrapolation of the time parameter $\tau_{int}/\tau_{y}$ to LHC energies. Widths and particle numbers denoted by * are extrapolated linearly with $\log(\sqrt{s_{NN}})$.
At RHIC energies the nonequilibrium sources  from quark-gluon interactions with particle content $N_{ch}^{1,2}$ dominate. At LHC energies the local equilibrium source from gluon-gluon collisions with particle content $N_{ch}^{eq}$ is the major origin of particle production at midrapidity.
Experimental midrapidity values (last column) are from PHOBOS \cite{bb01,bb03} for $|\eta| < 1$ at RHIC energies and from ALICE \cite{aamo10}  for $|\eta| < 0.5$ at
2.76 TeV.} 
\begin{tabular}{lllllllcr}
\hline\
$\sqrt{s_{NN}} $&$y_{beam}$& $\tau_{int}/\tau_{y}$&$<y_{1,2}>$&$\Gamma_{1,2}$&$\Gamma_{eq}$&$N_{ch}^{1,2}$&$N_{ch}^{eq}$&$\frac{dN}{d\eta}|_{\eta \simeq 0}$\\
   (TeV)\\
\hline
     
  0.13&$\mp 4.93$&0.89&$\mp 2.02$& 3.56&2.64&1837&560&547$\pm 55$\cite{bb01}\\
 0.20&$\mp 5.36$&0.80&$\mp 2.40$& 3.51&3.20&1887&1349&645$\pm 65$ \cite{bb03}\\
 2.76&$\mp 7.99$&0.67& $\mp 4.09$&4.2*&6.8*&3660*&11075&1601$\pm 60$ \cite{aamo10}\\
  5.52&$\mp 8.68$&0.66& $\mp 4.49$&4.6*&7.5*&4120*&14210*&1860*\\

\hline
\end{tabular}
\end{center}
\end{table}

\newpage
The dependencies of the diffusion-model parameters on incident energy, mass and centrality 
at RHIC energies have been investigated for various systems in \cite{wob06,kw07,kwo07}. In particular, the centrality dependence seen in the RHIC data is exactly reproduced \cite{wob06,kw07}. The parameters are shown in Fig.~\ref{fig2}  and Table~\ref{tab1} as functions of the c.m. energy in central collisions of Au + Au, and in an extrapolation to Pb + Pb at LHC energies. The difference between these two systems is very small since the diffusion-model parameters scale with the extension of the system like $A^{1/3}$, which differs only by a factor of 1.02. 

The time parameter
$\tau_{int}/\tau_{y}$ is displayed as function of center-of-mass energy in
the upper frame of Fig.~\ref{fig2}, with a functional dependence 
on the beam rapidity $y_{beam}$ and hence, on energy given by
$\tau_{int}/\tau_{y} \propto y_{beam}N_{part}/\sinh(y_{beam})$ 
as motivated in \cite{kwo07},
whereas the dashed curve assumes an exponential dependence
that yields a broader distribution function, see Fig.~8 in \cite{kwo07}
for a detailed comparison of the two limiting cases.
At LHC energies of 2.76 and 5.52 TeV for Pb + Pb I use in this prediction intermediate values between the two analytical extrapolations, see Table~\ref{tab1}.

The partial widths (FWHM) as functions of energy within the RHIC 
range for Au + Au are displayed in the 
middle and lower frames of Fig.~\ref{fig2} for both 
fragmentation and midrapidity
sources.
Here the widths are effective values: beyond the statistical widths
that can be calculated from a dissipation-fluctuation theorem \cite{wols99} within the RDM,
they include the effect of collective expansion.
The values at RHIC energies are resulting from a $\chi^{2}$-minimization with respect to the data that corresponds to the
time evolution up to $\tau_{int}$: The integration is 
stopped at the optimum values of 
$\tau_{int}/\tau_{y}$, $\Gamma_{1,2,eq}$, and $N_{ch}^{eq}$;
the explicit value of $\tau_{int}$ is not needed.

The normalization is given by the total number of produced charged hadrons that is taken from experiment if available, or extrapolated in case of predictions at higher energies. Hence, the model contains five parameters for symmetric systems, and six parameters for asymmetric systems. It 
provides an analytical framework to calculate the distribution function, and to draw physical conclusions.


The charged-particle distribution in rapidity space is obtained
as incoherent 
superposition of nonequilibrium and central (``equilibrium") solutions of
 (\ref{fpe}) 
\begin{eqnarray}
    \lefteqn{
\frac{dN_{ch}(y,t=\tau_{int})}{dy}=N_{ch}^{1}R_{1}(y,\tau_{int})}\nonumber\\&&
\qquad\qquad +N_{ch}^{2}R_{2}(y,\tau_{int})
+N_{ch}^{eq}R_{eq}(y,\tau_{int}).
\label{normloc1}
\end{eqnarray}



The results for pseudorapidity distributions of produced charged 
hadrons in central
Au + Au collisions at at two RHIC energies are shown in Fig.~\ref{fig3}
in comparison with PHOBOS data \cite{bb01,bb03}. In the $\chi^2-$minimization, the three-sources model yields excellent agreement with the data. Here the overall normalization is taken from the data, and the fit parameters are the time parameter (that determines the mean values $<y_{1,2}>$), the widths $\Gamma_{1,2}, \Gamma_{eq}$, and the number of produced particles in the central source $N_{ch}^{eq}$.

At RHIC energies, the multiplicity density at midrapidity has still a substantial contribution from the overlapping fragmentation sources. At 0.13 TeV, the contribution from the three sources at $\eta=0$ is about equal, at 0.2 TeV the midrapidity source is larger (58\%), but the fragmentation sources still contribute 21\% each.

It should be mentioned that there exist detailed microscopic calculations of fragmentation sources from 
$gq \rightarrow q$ and $qg \rightarrow q$ diagrams by Szczurek et al. \cite{sz04,csz05} for pion production in proton-proton, and nucleus-nucleus collisions at SPS and RHIC energies. These processes are also responsible for the observed differences \cite{bea01} in the production of positively and negatively charged hadrons, in particular, pions. An extension of these calculations to LHC energies is very desirable.

Within the 3-sources RDM, we had presented predictions at LHC energies of 5.52 TeV in \cite{kwo07} that were included in \cite{arm08,arme09}. The total number of produced charged hadrons had been extrapolated with $\log(\sqrt{s_{NN}})$ to obtain $26.5*N_{part}$ at 5.52 TeV, with the number of participants $N_{part}$. Based on this assumption, the calculated RDM-pseudorapidity distribution function turned out to underestimate the midrapidity result that is expected using the recent ALICE 2.76 TeV data point \cite{aamo10} by a factor of 2.7.

I have now chosen to adjust the RDM
parameters such that the ALICE midrapidity value at 2.76 TeV is reproduced, 
1601$\pm60$ 
\cite{aamo10} (1584$\pm4~ $(stat.)$\pm76~ $    (sys.) in \cite{aa10}). On this basis, the RDM distribution functions at 2.76 and 5.52 TeV can be calculated. 

With the extrapolation of the time parameter and the partial widths $\Gamma_{1,2,eq}$ from Fig.~\ref{fig2}, plus corresponding extrapolations of the number of produced particles in fragmentation and central sources as functions of $\log(\sqrt{s_{NN}})$ given in Table~\ref{tab1}, the results are shown in Fig.~\ref{fig3}. The main uncertainty is in the extrapolation of the particle content of the fragmentation sources since the content of the central source is essentially fixed by the ALICE midrapidity data point. The calculation at 5.52 TeV is performed based on an extrapolation of the multiplicity density at midrapidity with $\log{(\sqrt{s_{NN}})}$ that yields $dN/d\eta \simeq 1860$ at midrapidity. 

At LHC energies, the overall scenario changes even more in favor of particle production from the midrapidity source. The bulk of the midrapidity density is generated in the central source (93\%),
there is only a small overlap of the fragmentation sources at midrapidity as shown in Fig.~\ref{fig4}.

In a comparison with calculations at LHC energies that do not include the Jacobian transformation as displayed by the dotted curves in Fig.~\ref{fig3},\ref{fig4}, it is evident that the midrapidity dip structure is essentially determined by the interplay of the three sources for particle production, and only marginally influenced by the transformation from y$-$ to $\eta-$space at these high energies. The central distribution including the Jacobian has no dip at LHC energies, but only a slight reduction in absolute magnitude at midrapidity, as shown by the dashed curve in Fig.~\ref{fig4}. 

The smallness of the fragmentation sources at midrapidity is in qualitative agreement with  results of a  microscopic model that we had developed in \cite{mtw09} to investigate net-baryon distributions at LHC energies. In that approach, the net-baryon yield at large rapidities is calculated from the interaction of valence quarks with the gluon condensate in the respective other nucleus. Extending the model to the midrapidity region \cite{mtw10}, a net-baryon midrapidity density dN/dy(y=0)$\simeq$4 is obtained at 5.52 TeV, corresponding to
a midrapidity density of 12 valence quarks -- as opposed to a total of 1248 valence quarks in the system. Hence the charged-hadron production from valence quark -- gluon interactions at LHC energies can be expected to be very small in the midrapidity region.
\newpage
\section{Conclusion}
Based on the description of charged-hadron pseudorapidity distributions
in central collisions of heavy symmetric systems at RHIC energies in a non-equilibrium-statistical model,
I have presented predictions of pseudorapidity distributions of produced charged hadrons for central Pb + Pb  collisions at LHC energies of 2.76 and 5.52 TeV.
These rely on the extrapolation of the transport parameters in the relativistic diffusion model (RDM) with increasing center-of-mass energy.

In a three-sources model, the midrapidity source that is associated with gluon-gluon collisions accounts for about 93\% of the charged-particle multiplicity density measured by ALICE at midrapidity in Pb + Pb collisions at 2.76 TeV. The fragmentation sources that correspond to particles that are mainly generated from valence quark -- gluon interactions are centered at relatively large values of pseudorapidity $(<\eta_{1,2}>\simeq<y_{1,2}> \simeq \mp 4.1)$ and hence, these contribute only marginally to the midrapidity yield.

Since the Jacobian transformation from rapidity to pseudorapidity space is close to 1 at LHC energies due to the large mean transverse momenta, the size of the midrapidity-dip in the pseudorapidity distribution function is essentially determined by the relative particle content in the three sources, not by the Jacobian. Small corrections of the extrapolated values for the number of produced particles in the fragmentation sources may be required once the measured distributions become available from CMS, ATLAS and ALICE at both LHC energies.\\
\newpage

\bf{Acknowledgments}

\rm
This work has been supported
by 
the ExtreMe Matter Institute EMMI.
\bibliographystyle{elsarticle-num}
\bibliography{gw_plb_nt}

\begin{thebibliography}{10}
\expandafter\ifx\csname url\endcsname\relax
  \def\url#1{\texttt{#1}}\fi
\expandafter\ifx\csname urlprefix\endcsname\relax\def\urlprefix{URL }\fi
\expandafter\ifx\csname href\endcsname\relax
  \def\href#1#2{#2} \def\path#1{#1}\fi

\bibitem{aa10}
K.~{A}amodt~{\it et al.} (ALICE~Collaboration), Phys. Rev. Lett. 105 (2010)
  252301--1--11.

\bibitem{aamo10}
K.~{A}amodt~{\it et al.} (ALICE~Collaboration), Phys. Rev. Lett. 106 (2011)
  032301--1--10.

\bibitem{arm08}
N.~{A}rmesto {\it et al.}, J. Phys. G 35 (2008) 054001--1--170.

\bibitem{arme09}
N.~Armesto, arXiv: 0903.1330.

\bibitem{wol99}
G.~Wolschin, Eur. Phys. J. A 5 (1999) 85--90.

\bibitem{wol07}
G.~Wolschin, Prog. Part. Nucl. Phys. 59 (2007) 374--382.

\bibitem{biy04}
M.~Biyajima, M.~Ide, M.~Kaneyama, T.~Mizoguchi, N.~Suzuki, Prog. Theor. Phys.
  Suppl. 153 (2004) 344--348.

\bibitem{wob06}
G.~Wolschin, M.~Biyajima, T.~Mizoguchi, N.~Suzuki, Annalen Phys. 15~(6) (2006)
  369--378.

\bibitem{kw07}
R.~Kuiper, G.~Wolschin, Europhys. Lett. 78 (2007) 2201--1--5.

\bibitem{wobi06}
G.~Wolschin, M.~Biyajima, T.~Mizoguchi, N.~Suzuki, Phys. Lett. B633 (2006)
  38--42.

\bibitem{bia05}
A.~Bialas, W.~Czyz, Acta Phys. Polon. B 36 (2005) 905--918.

\bibitem{cap82}
A.~Capella, J.~Kwieci\'nski, J.~T.~T. Van, Phys. Lett. 108B (1982) 347--350.

\bibitem{cap94}
A.~{C}apella {\it et al.}, Phys. Rept. 236 (1994) 225--329.

\bibitem{kai03}
A.~Kaidalov, Yad. Fiz. 66 (2003) 2044--2066.

\bibitem{bb04}
B.~B. {B}ack~{\it et al.} (PHOBOS~Collaboration), Phys. Rev. Lett. 93 (2004)
  082301--1--4.

\bibitem{bb05}
B.~B. {B}ack~{\it et al.} (PHOBOS~Collaboration), Phys. Rev. C 72 (2005)
  0319018(R)--1--5.

\bibitem{alv11}
B.~{A}lver~{\it et al.} (PHOBOS~Collaboration), Phys.Rev.C 83 (2011)
  024913--1--24.

\bibitem{alb00}
W.~M. Alberico, A.~Lavagno, P.~Quarati, Eur. Phys. J. C 12 (2000) 499--506.

\bibitem{ryb03}
M.~Rybczy\'nski, Z.~W{\l}odarczyk, G.~Wilk, Nucl. Phys. B (Proc. Suppl.) 122
  (2003) 325--328.

\bibitem{uhl30}
G.~Uhlenbeck, L.~Ornstein, Phys. Rev. 36 (1930) 823--841.

\bibitem{wol03}
G.~Wolschin, Phys. Lett. B569 (2003) 67--72.

\bibitem{bb01}
B.~B. {B}ack~{\it et al.} (PHOBOS~Collaboration), Phys. Rev. Lett. 87 (2001)
  102303--1--4.

\bibitem{bb03}
B.~B. {B}ack~{\it et al.} (PHOBOS~Collaboration), Phys. Rev. Lett. 91 (2003)
  052303--1--4.

\bibitem{kwo07}
R.~Kuiper, G.~Wolschin, Annalen Phys. 16~(1) (2007) 67--77.

\bibitem{wols99}
G.~Wolschin, Europhys. Lett. 47 (1999) 30--35.

\bibitem{sz04}
A.~Szczurek, Acta Phys. Polon. B 35 (2004) 161--168.

\bibitem{csz05}
M.~Czech, A.~Szczurek, Phys. Rev. C 72 (2005) 015202--1--11.

\bibitem{bea01}
I.~G. {B}earden~{\it et al.} (BRAHMS~Collaboration), Phys. Rev. Lett. 87 (2001)
  112305--1--4.

\bibitem{mtw09}
Y.~Mehtar-Tani, G.~Wolschin, Phys. Rev. Lett. 102 (2009) 182301--1--4.

\bibitem{mtw10}
Y.~Mehtar-Tani, G.~Wolschin, Phys. Lett. B 688 (2010) 174--177.

\end{thebibliography}


\end{document}